\documentclass[aps,prl,reprint,groupedaddress]{revtex4-1}

\usepackage{graphicx}
\usepackage{amssymb}

\begin{document}

%\preprint{}

\title{Comparison of model predictions for elliptic flow with experiment for $Pb+Pb$ collisions at $\sqrt {s_{NN}}=2.76$ TeV}


\author{T. J. Humanic}
\email[]{humanic@mps.ohio-state.edu}
%\thanks{}
\affiliation{Department of Physics, The Ohio State University,
Columbus, Ohio, USA}


\date{\today}

\begin{abstract}
A simple kinematic model based on the superposition of $p+p$ collisions, relativistic 
geometry and hadronic rescattering
is used to predict the elliptic flow observable in $\sqrt {s_{NN}} = 2.76$ TeV $Pb+Pb$ collisions. 
A short proper time for hadronization is assumed. The predictions are compared with
recent experimental measurements of elliptic flow which have been made for this
colliding system and energy. It is found that the model predictions do a reasonable
job in describing the experimental results, suggesting that the parton phase in these
collisions may be short-lived.
\end{abstract}

%\keywords{}

\maketitle

The CERN Large Hadron Collider has recently begun delivering $Pb+Pb$ collisions
at $\sqrt{s_{NN}}=2.76$ TeV to experiments. These are
 the highest energy heavy-ion collisions ever to be
produced in the laboratory. The LHC ALICE experiment\cite{Aamodt:2008zz} has already posted two
experimental papers based on data from these collisions, one in which the charged
particle (hadron) multiplicity density is measured\cite{Aamodt:2010pb} and another 
in which the elliptic flow
of charged hadrons is measured\cite{Aamodt:2010pa}. These are both important and basic
observables to measure in heavy-ion collisions since the charged particle multiplicity 
density is related to the
initial particle density and the elliptic flow is sensitive to the initial dynamics of the particles, both of which could be messengers of possible exotic phenomena taking place in these collisions\cite{Aamodt:2008zz}.

In order to better understand the underlying physics driving these observables at the LHC, a simple kinematic model 
has been constructed\cite{Humanic:2010su,Humanic:2008nt}
with the goal of comparing predictions of this model with the experimentally measured observables.
The basis of the model is that the initial state of the heavy-ion collision is determined by the superposition of proton-proton collisions followed by the mutual scattering of the hadrons produced in the collision.
Besides its simplicity, the advantages of this 
model are 1) the model
has been shown to describe the overall trends of hadronic observables in lower energy
$Au+Au$ collisions at $\sqrt{s_{NN}}=$ 0.20 TeV from the Relativistic Heavy Ion Collider 
(RHIC)\cite{Humanic:2008nt}, and 2) the model is easily scalable to LHC energies.
These will be ``limiting case scenario'' predictions in the sense that only hadrons are
used as the degrees of freedom in this model even at the early stages of the
collision where parton (quark and gluon) degrees of freedom are thought to be
more appropriate, i.e. a short
proper time for hadronization is assumed.
In spite of this assumption, it is interesting to note that at RHIC energies, 
this model has even been able to satisfactorily reproduce the quark number
scaling seen in experiment for elliptic flow measurements for 
$Au+Au$ collisions\cite{Humanic:2008nt}. Quark number scaling seen in
experimental observables at RHIC is often cited as evidence that Quark Matter is produced
in RHIC collisions\cite{Adams:2004bi}

A prediction of the multiplicity density from this model has 
already been compared with the measured
multiplicity density in central $Pb+Pb$ collisions at $\sqrt{s_{NN}}=2.76$ TeV from ALICE
in Reference \cite{Aamodt:2010pb}.
The outcome is that the model under-predicts the multiplicity density observable, $dN/d\eta$,
at mid-rapidity by $\sim 20\%$ with respect to the lower bound of the measurement.
Considering that the model was used to extrapolate from RHIC to LHC energy
which represents a factor of 14 increase in $\sqrt{s_{NN}}$,
it is somewhat encouraging that  the model prediction is even this close to the measurement, and could be considered a validation to some degree of the simple approach taken in the model.

In this paper a further and more stringent test of the model is presented by comparing model predictions 
for elliptic flow with the recent ALICE measurements of this observable\cite{Aamodt:2010pa}. The model
predictions are based on the same calculations which were made in the prediction of the
multiplicity density mentioned above in order to get a consistent picture of the ability of this simple model
to describe the experimental results. A brief description of the model is presented below
followed by the comparison of the model predictions for elliptic flow with the ALICE experiment
for $Pb+Pb$ collisions at $\sqrt{s_{NN}}=2.76$ TeV.

The model calculations are carried out in five main steps: 1) generate hadrons in $p+p$  collisions from the event-generator PYTHIA, 2) superpose $p+p$ collisions in the geometry of the colliding nuclei, 3) employ a simple space-time geometry picture for the hadronization of the
PYTHIA-generated hadrons,   4) calculate the effects of rescattering among the hadrons,
and 5) calculate the hadronic observables. These steps will now be discussed in more detail.

The $p+p$ collisions were modeled with the PYTHIA code \cite{pythia6.4}, version 6.409. The internal parton distribution functions ``CTEQ 5L'' (leading order) were used in these calculations. Events were generated
in ``minimum bias'' mode, i.e. setting the low-$p_T$ cutoff for parton-parton collisions to zero (or
in terms of the actual PYTHIA parameter, $ckin(3)=0$) and excluding elastic and diffractive collisions (PYTHIA parameter $msel=1$). Runs were made at $\sqrt{s}= $ 2.76 TeV to simulate 
LHC collisions. Information saved
from a PYTHIA run for use in the next step of the procedure were the momenta and identities
of the ``direct'' (i.e. redundancies removed) hadrons (all charge states) $\pi$, $K$, $p$, $n$,
$\Delta$, $\Lambda$, $\rho$, $\omega$, $\eta$, ${\eta}'$, $\phi$, and $K^*$. These particles were
chosen since they are the most common hadrons produced and thus should have the greatest
effect on the hadronic observables in these calculations.

A main assumption of the model is that an adequate job of describing the heavy-ion collision can be obtained by superposing PYTHIA-generated  $p+p$ collisions calculated at the beam $\sqrt s$
within the collision geometry of the colliding nuclei. Specifically, for a collision of impact
parameter $b$, if $f(b)$ is the fraction of the overlap volume of the participating parts of the nuclei such that $f(b=0)=1$ and $f(b=2R)=0$, where $R=1.2A^{1/3}$ and $A$ is the mass 
number of the nuclei, then
the number of $p+p$ collisions to be superposed will be $f(b)A$. The positions of the superposed $p+p$ pairs are randomly distributed in the overlap volume and then projected onto the $x-y$ plane which is transverse to the beam axis defined in the $z$-direction. The coordinates for a particular
$p+p$ pair are defined as $x_{pp}$, $y_{pp}$, and $z_{pp} = 0$. 
The positions of the hadrons produced in 
one of these $p+p$ collisions are defined with respect to the position so obtained of the superposed
$p+p$ collision (see below). 

As was done in similar calculations for lower-energy RHIC collisions to give better 
agreement with experimental $dn/d\eta$ distributions\cite{Humanic:2008nt}, a lower multiplicity cut
was applied to the $p+p$ collisions used in the present calculations which 
rejected the lowest ~20\% of the events. The spirit of this cut is to partially compensate for the fact that there is no re-interaction of primary nucleons from the projectile-target system in this model.

The space-time geometry picture for hadronization from a superposed $p+p$
collision located at $(x_{pp},y_{pp})$ consists of the emission of a PYTHIA
particle from a thin uniform disk of radius 1 fm in the $x-y$ plane followed by
its hadronization which occurs in the proper time of the particle, $\tau$. The space-time
coordinates at hadronization in the lab frame $(x_h, y_h, z_h, t_h)$ for a particle with momentum
coordinates $(p_x, p_y, p_z)$, energy $E$, rest mass $m_0$, and transverse disk
coordinates $(x_0, y_0)$, which are chosen randomly on the disk,  can then be written as

\begin{eqnarray}
x_h = x_{pp} + x_0 + \tau \frac{p_x}{m_0} \\
y_h = y_{pp} + y_0 + \tau \frac{p_y}{m_0} \\
z_h = \tau \frac{p_z}{m_0} \\
t_h = \tau \frac{E}{m_0}
\end{eqnarray}

The simplicity of this geometric picture is now clear: it is just an expression of causality with the
assumption that all particles hadronize with the same proper time, $\tau$. A similar hadronization
picture (with an initial point source) has been applied to $e^+-e^-$ collisions\cite{csorgo}.
For all results presented in this work,  $\tau$ will be set to 0.1 fm/c as was done in
applying the present model to calculating predictions for 
RHIC $Au+Au$ collisions\cite{Humanic:2008nt} and 
Tevatron $p+\bar{p}$ collisions\cite{Humanic:2006ib}.

The hadronic rescattering calculational method used is similar to that
employed in previous studies \cite{Humanic:1998a,Humanic:2006a}.
Rescattering is simulated with a semi-classical Monte Carlo
calculation which assumes strong binary collisions between hadrons.
Relativistic kinematics is used throughout. The hadrons considered in the
calculation are the most common ones: pions, kaons,
nucleons and lambdas ($\pi$, K,
N, and $\Lambda$), and the $\rho$, $\omega$, $\eta$, ${\eta}'$,
$\phi$, $\Delta$, and $K^*$ resonances.
For simplicity, the
calculation is isospin averaged (e.g. no distinction is made among a
$\pi^{+}$, $\pi^0$, and $\pi^{-}$).

The rescattering calculation finishes
with the freeze out and decay of all particles. Starting from the
initial stage ($t=0$ fm/c), the positions of all particles in each event are
allowed to evolve in time in small time steps ($\Delta t=0.5$ fm/c)
according to their initial momenta. At each time step each particle
is checked to see a) if it has hadronized ($t>t_h$, where $t_h$ is given in
Eq. (4)), b) if it
decays, and c) if it is sufficiently close to another particle to
scatter with it. Isospin-averaged s-wave and p-wave cross sections
for meson scattering are obtained from Prakash et al.\cite{Prakash:1993a}
and other cross sections are estimated from fits to hadron scattering data
in the Review of Particle Physics\cite{pdg}. Both elastic and inelastic collisions are
included. The calculation is carried out to 400 fm/c which
allows enough time for the rescattering to finish (as a test, calculations were also carried out for
longer times with no changes in the results). Note that when this cutoff time is reached, all un-decayed resonances are allowed to decay with their natural lifetimes and their projected decay positions and times are recorded.

The rescattering calculation is described in more detail elsewhere
\cite{Humanic:2006a,Humanic:1998a}. The validity of the numerical
methods used in the rescattering code have been studied using
the subdivision method, the results of which have verified that the methods used are 
valid \cite{Humanic:2006b}.

Model runs are made to be ``minimum bias'' by having the impact parameters of collisions follow the distribution $d\sigma/db \propto  b$, where $0<b<2R$. Observables are then calculated from the model in the appropriate centrality bin by making multiplicity cuts as normally done in experiments, as well as kinematic cuts on rapidity and $p_T$. For the present study, a 3200 event minimum bias run was made from the model for $\sqrt {s_{NN}}=2.76$ TeV $Pb+Pb$ collisions which was then used to calculate hadronic observables. As mentioned earlier, these are the same events which were used in the
prediction of the multiplicity density for Reference \cite{Aamodt:2010pb}.

%\subsection{}
%\subsubsection{}

% If in two-column mode, this environment will change to single-column
% format so that long equations can be displayed. Use
%\begin{widetext}
%\end{widetext}

% figures should be put into the text as floats.
% Use the graphics or graphicx packages (distributed with LaTeX2e)
% and the \includegraphics macro defined in those packages.
% See the LaTeX Graphics Companion by Michel Goosens, Sebastian Rahtz,
% entire page. There is no need to do explicit centering.

% \begin{figure}
% \end{figure}

\begin{figure*}
\begin{center}
\includegraphics[width=160mm]{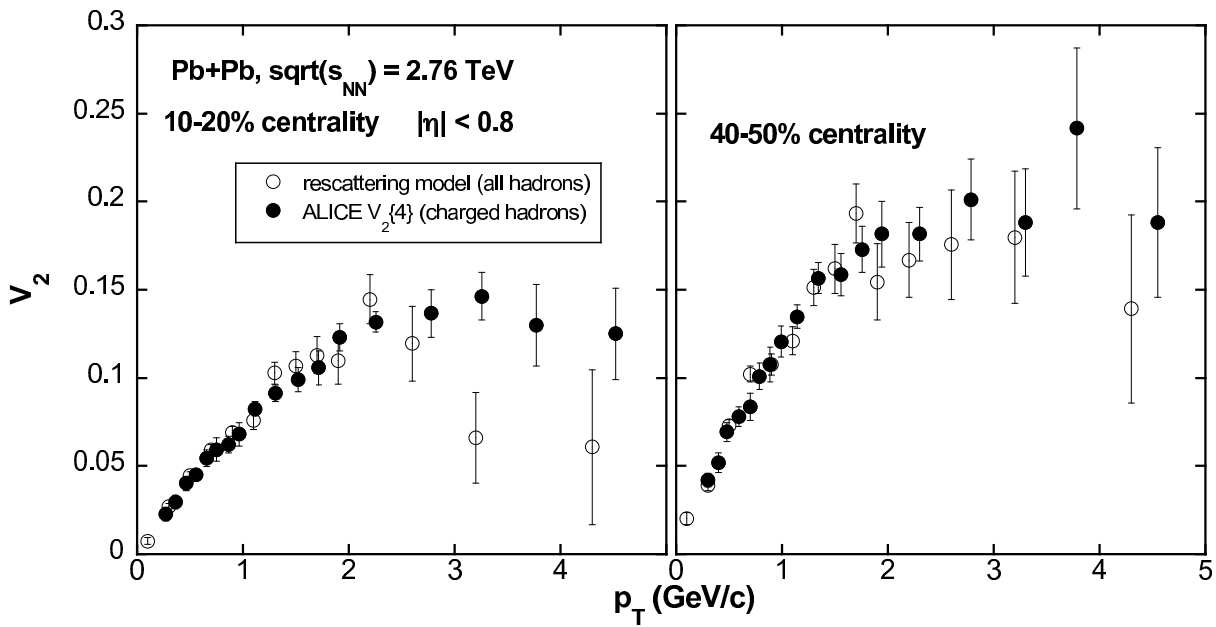} \caption{Comparison of model $V_2$ vs. $p_T$ plots for
$\sqrt{s_{NN}}=2.76$ TeV $Pb+Pb$ collisions with the ALICE experiment
for $10-20\%$ and $40-50\%$ centrality classes.}
\label{fig1}
\end{center}
\end{figure*}

The elliptic flow variable, $V_2$, is
defined as
\begin{eqnarray}
\label{v2} V_2=<\cos(2\phi)> \\\nonumber
    \phi=\arctan(\frac{p_y}{p_x})
\end{eqnarray}
where ``$<>$'' implies a sum over particles in an event and a sum over events
and where $p_x$ and $p_y$ are the $x$ and $y$ components of the particle
momentum, and $x$ is in the impact parameter direction, i.e.
reaction plane direction, and $y$ is in the direction perpendicular
to the reaction plane. The $V_2$ variable is calculated from the model
using Eq. (\ref{v2}) and taking the reaction plane to be the model $x-z$ plane.
As seen, if $<p_x> \sim <p_y>$, then $V_2 \sim 0$, and for $<p_x> \gg <p_y>$,
then $V_2\sim 1$.

Figure \ref{fig1} shows the comparison of model $V_2$ vs. $p_T$ plots for $\sqrt{s_{NN}}=2.76$ TeV $Pb+Pb$ collisions with measurements from the ALICE experiment\cite{Aamodt:2010pa} 
for $10-20\%$ and $40-50\%$ centrality classes. Note that all model predictions in this
paper are compared with the experimental $V_2\{4\}$ value, which is $V_2$ extracted using
the 4-particle cumulant method\cite{Aamodt:2010pa}, since this observable minimizes 
non-flow correlations in the elliptic flow which are not present in the model, thus making
it more comparable. The model predictions are seen to follow the experimental
points fairly closely within the uncertainties shown for both centrality classes. The uncertainties
on the model prediction are statistical. The model accurately describes the increasing $V_2$
with increasing $p_T$ for $p_T<2$ GeV/c and then the ``flattening'' of the dependence of
$V_2$ on further increase of $p_T$ for $p_T> 2$ GeV/c, although somewhat under-predicting
the measurement for $p_T>2$ GeV/c for the $10-20\%$ centrality class.
From the
model picture all of these dependences, as well as the non-vanishing values of $V_2$, are
a result of the  hadronic rescattering. If rescattering is turned off in the model,
or equivalently $\tau \gg 0.1$ fm/c,
$V_2\rightarrow 0$ for all cases\cite{Humanic:2010su,Humanic:2008nt}.

\begin{figure}
\begin{center}
\includegraphics[width=85mm]{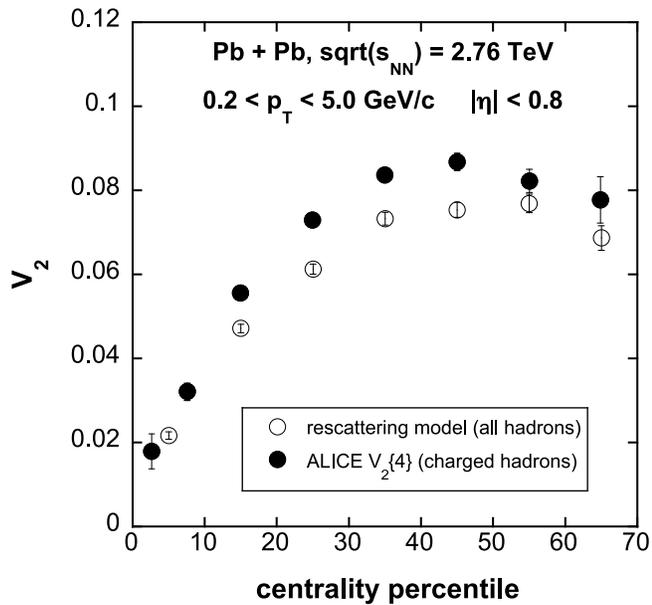} \caption{Comparison of model integrated $V_2$ vs.
centrality class for $\sqrt{s_{NN}}=2.76$ TeV $Pb+Pb$ collisions with the ALICE experiment.}
\label{fig2}
\end{center}
\end{figure}

Figure \ref{fig2} compares $V_2$ integrated over $0.2<p_T<5.0$ GeV/c vs. centrality class
with the ALICE experiment\cite{Aamodt:2010pa}. The model is seen to qualitatively 
describe the trend of the measurements
of increasing $V_2$ with increasing centrality class and then reaching a maximum,
but to under-predict the measurements on average by $\sim 14\%$. This under-prediction
could be a reflection of the under-prediction of the model for the particle multiplicity, mentioned
earlier (which could be corrected in the model by including some degree of re-interaction
of the primary nucleons from the projectile-target system).

From the comparisons shown above in Figures \ref{fig1} and \ref{fig2} between the model and experiment, it is seen that the model does a reasonably good job in describing the main features
of the experimental elliptic flow results for LHC $Pb+Pb$ collisions. 
The model even shows the ``high-$p_T$ flattening'' feature of $V_2$ which pure
hydrodynamic models that are based on the formation of Quark Matter in these
collisions and thus use parton degrees of freedom in the early stage of the collision are not able to reproduce \cite{Kestin:2009}.
While it is interesting
and perhaps a little disturbing that so simple a model is able to do as well as it does in describing
the experimental results, the question of ``what physics have we learned from this?'' could be posed.
The main motivation for this model has been to study to what extent hadronic observables in
relativistic heavy-ion collisions are influenced by  hadronic rescattering.
It is expected that some hadronic rescattering will take place in these collisions after some
transition from a possible Quark Matter state to hadronic matter.
As mentioned
earlier, the ``limiting case scenario'' has been studied in the present model where only 
hadronic degrees of freedom are considered. An initial-state model is used
to set
up the initial kinematics, in the present case a superposition of PYTHIA $p+p$ collisions due to its convenience
but which alternatively could have been a thermal or parton model as have been used in a past successful
elliptic flow model study
for RHIC collisions \cite{Boggild:2009}, but the hadronic rescattering is the ``active ingredient'' driving the dynamics of the hadronic observables in the model. In order to get enough rescattering to agree with
experiment, the rescattering must begin almost immediately in the collision. This suggests three possible interpretations
as to the meaning of the model at the early stage of the collision where 
its assumptions are the most contentious: 1) the hadronization time is
very short, i.e. $\tau=0.1$ fm/c, and thus hadronic degrees of freedom dominate from the beginning, 
2) the early stage is partonic and more extended in
time and the early-stage hadronic scattering in the model ``mocks up'' some qualitative features of
the early-stage parton scattering (including ``viscosity'' in the language of hydrodynamic models),
or 3) a combination of 1) and 2) wherein a mixed phase of partons and hadron-like objects
initially coexist until complete hadronization occurs.
Further model studies and comparisons with experiment for other observables may help 
clarify which of these interpretations
is the more valid.

In conclusion, a kinematic model based on the superposition of PYTHIA-generated $p+p$ collisions, relativistic geometry and  hadronic rescattering
has been used in the present work to predict the elliptic flow observable in $\sqrt {s_{NN}} = 2.76$ TeV $Pb+Pb$ collisions. A short proper time for hadronization of $\tau=0.1$ fm/c
has been assumed as in previous studies with this model which have shown qualitative
agreement with experiments. It has been shown that the simple hadronic rescattering model
accurately describes the features of the ALICE elliptic flow measurements,
suggesting that hadronic rescattering plays an important role in determining the
properties of the elliptic flow observable in these collisions. These results also suggest
that the parton phase in these collisions may be short-lived.

%\begin{acknowledgments}
%\end{acknowledgments}


\begin{acknowledgments}
The author wishes to acknowledge financial support from the U.S.
National Science Foundation under grant PHY-0970048, and to acknowledge computing
support from the Ohio Supercomputing Center.
\end{acknowledgments}

\end{document}